\documentclass[pre,twocolumn,aps,showpacs]{revtex4}
\newcommand{\bec}[1]{\mbox{\boldmath $ #1$}}
\begin{document}
\centerline{\bf PHYSICAL REVIEW E, v. 59, 6724-6729 (1999)}
\bigskip
\bigskip
\title{Magnetic helicity tensor for an anisotropic turbulence}
\author{Nathan Kleeorin}
\email{nat@menix.bgu.ac.il}
\author{Igor Rogachevskii}
\email{gary@menix.bgu.ac.il} \homepage{http://www.bgu.ac.il/~gary}
\affiliation{Department of Mechanical Engineering, The Ben-Gurion
University of the Negev, \\
POB 653, Beer-Sheva 84105, Israel}
\date{Received 28 September 1998}
\begin{abstract}
The evolution of the magnetic helicity tensor for a nonzero mean
magnetic field and for large magnetic
Reynolds numbers in an anisotropic turbulence is studied. It is shown that the
isotropic and anisotropic parts of the magnetic helicity tensor
have different characteristic times of evolution. The time of
variation of the isotropic part of the magnetic helicity tensor
is much larger than the correlation time of the turbulent velocity field.
The anisotropic part of the magnetic helicity tensor
changes for the correlation time of the turbulent velocity field.
The mean turbulent flux of the magnetic
helicity is calculated as well. It is shown that even a small anisotropy
of turbulence strongly modifies the flux of the magnetic helicity.
It is demonstrated that the tensor of the magnetic part of the
$ \alpha $-effect for weakly inhomogeneous
turbulence is determined only by the isotropic part of the magnetic
helicity tensor.
\end{abstract}
\maketitle

\section{Introduction}

The magnetic helicity $ {\bf A}^{(t)} \cdot {\bf H} $ is a fundamental
quantity in magnetohydrodynamics because it is
conserved in the limit of infinite electrical conductivity of the
medium, where $ {\bf H} = \bec{\bf \nabla} \times {\bf A}^{(t)} $
is the magnetic field and $ {\bf A}^{(t)} $
is the magnetic vector potential. In addition, the topological properties
of magnetic field are determined by the magnetic helicity (see, e.g.,
\cite{M78,ZRS83}). In  developed magnetohydrodynamic turbulence
the mean magnetic helicity
$ \langle {\bf a} \cdot {\bf h} \rangle $ is conserved as well in the limit of
infinite magnetic Reynolds numbers and zero mean magnetic field,
where $ {\bf h} $ and $ {\bf a} $
are fluctuations of the magnetic field and the magnetic vector
potential, respectively (see, e.g., \cite{M78,ZRS83}).
The magnetic helicity tensor $ \chi_{ij} =
\langle a_{i}({\bf x}) h_{j}({\bf x}) \rangle $  determines
the tensor of the magnetic part of the $ \alpha $-effect.
The latter is of fundamental importance in view of magnetic dynamo
(see, e.g., \cite{M78,ZRS83,KR80}).
In spite of the great importance of this quantity, a dynamics of the
magnetic helicity tensor for an anisotropic turbulence is poorly
understood.

In the present paper the equation for the magnetic helicity tensor
for an anisotropic turbulence and a nonzero mean magnetic field, and
for large magnetic Reynolds numbers is derived. It is shown that the
isotropic and anisotropic parts of the magnetic helicity tensor
have different characteristic times of evolution. The time of
variation of the isotropic part of the magnetic helicity tensor
is much longer than the correlation time of the turbulent velocity field.
On the other hand, the anisotropic part of the magnetic helicity tensor
changes for the correlation time of the turbulent velocity field.
This anisotropic part is determined only by the turbulent magnetic
diffusion tensor. The mean turbulent flux of the magnetic helicity
is calculated as well. It is shown that even small anisotropy of turbulence
strongly modifies the flux of the magnetic helicity.

\section{The equation for the magnetic helicity: simple approach}

First, we derive an equation for the magnetic helicity
for an anisotropic turbulence by a simple
consideration. The induction equation for the magnetic field $ {\bf H} $
is given by
\begin{eqnarray}
\partial {\bf H} / \partial t = \bec{\bf \nabla} \times ({\bf v}
\times {\bf H}  - \eta \bec{\bf \nabla} \times {\bf H}) \;,
\label{A1}
\end{eqnarray}
where $ {\bf H} = {\bf B} + {\bf h} ,$ and $ {\bf B} = \langle
{\bf H} \rangle $ is the mean magnetic field,
$ {\bf v} = {\bf V} + {\bf u} ,$ and $ {\bf V} = \langle
{\bf v} \rangle $ is the mean fluid velocity field, $ \eta $ is the
magnetic diffusion due to electrical conductivity of fluid.
The equation for the vector potential $ {\bf A}^{(t)} $ follows
from the induction equation (\ref{A1})
\begin{eqnarray}
\partial {\bf A}^{(t)} / \partial t = {\bf v} \times {\bf H}  -
\eta \bec{\bf \nabla} \times (\bec{\bf \nabla} \times {\bf A}^{(t)}) +
\bec{\bf \nabla} \varphi \;,
\label{A2}
\end{eqnarray}
where $ {\bf H} = \bec{\bf \nabla} \times {\bf A}^{(t)} ,$ and
$ {\bf A}^{(t)} = {\bf A} + {\bf a} ,$ and $ {\bf A} = \langle
{\bf A}^{(t)} \rangle $ is the mean vector potential,
$ \varphi = \tilde \Phi + \phi $ is an arbitrary scalar function,
and $ \tilde \Phi =  \langle \varphi \rangle .$ Now we multiply Eq.
(\ref{A1}) by $ {\bf a} $ and Eq. (\ref{A2}) by $ {\bf h} $, add them and
average over the ensemble of turbulent fields. This yields an equation
for the magnetic helicity $ \chi = \langle a_p({\bf x})
h_p({\bf x}) \rangle :$
\begin{eqnarray}
\partial \chi / \partial t = - 2 \langle {\bf u} \times {\bf
h} \rangle \cdot {\bf B} - 2 \eta \langle {\bf h} \cdot (\bec{\bf \nabla}
\times {\bf h}) \rangle - \bec{\bf \nabla} \cdot \tilde {\bf F} \;,
\label{A3}
\end{eqnarray}
where $ \tilde F_{p} = V_{p} \chi - \chi_{pn} V_{n}+ \langle {\bf a} \times
{\bf u} \rangle \times {\bf B} - \eta \langle {\bf a} \times (\bec{\bf \nabla}
\times {\bf h}) \rangle + \langle {\bf a} \times ({\bf u} \times {\bf h} )
\rangle - \langle {\bf h} \phi \rangle .$
Electromotive force for an anisotropic turbulence is given by
\begin{eqnarray}
\langle {\bf u \times h} \rangle =
{\bf V}_{\rm DM} {\bf \times } {\bf B}
+ \hat \alpha {\bf B} - \hat \eta \bec{\bf \nabla} {\bf \times} {\bf B} \;
\label{A4}
\end{eqnarray}
(see, e.g., \cite{R80,RK97}), where $ \hat \eta \equiv \hat \eta_{mn}
= (\eta_{pp} \delta_{mn} - \eta_{mn}) / 2 ,$ and $ \eta_{mn} =
\eta \delta_{mn} + \tilde \eta_{mn} ,$ and $ \tilde \eta_{mn} =
\langle \tau u_m u_n \rangle ,$ and
$ ({\bf V}_{\rm DM})_{n} = - {\bf \nabla}_{m} \tilde \eta_{mn} / 2 $ is the
velocity caused by the turbulent diamagnetism, and $ \hat \alpha =
\alpha_{mn} = \alpha_{mn}^{(v)} + \alpha_{mn}^{(B)} ,$ and
the tensors $ \alpha^{(v)}_{mn} $ and $ \alpha^{(B)}_{mn} $
are given by
\begin{eqnarray}
\alpha^{(v)}_{mn} &=& - [\varepsilon_{mji} \langle \tau u_i({\bf
x}) {\bf  \nabla}_{n} u_j ({\bf x}) \rangle
\nonumber\\
&& + \varepsilon_{nji} \langle \tau u_i ({\bf x}) {\bf \nabla}_{m}
u_j ({\bf x}) \rangle ] / 2 \;,
\label{A5} \\
\alpha^{(B)}_{mn} &=&  [\varepsilon_{mji} \langle \tau h_i ({\bf x})
{\bf  \nabla}_{n} h_j ({\bf x}) \rangle
\nonumber\\
&& + \varepsilon_{nji} \langle \tau h_i ({\bf x}) {\bf
\nabla}_{m} h_j ({\bf x}) \rangle] / (2 \mu_{0} \rho) \; .
\label{A6}
\end{eqnarray}
Substituting Eq. (\ref{A4}) into Eq. (\ref{A3}) we obtain after
simple manipulations an equation for the magnetic helicity:
\begin{eqnarray}
{\partial \chi \over \partial t} &=& - 2 \eta \biggl({\partial^{2}
\chi \over
\partial x_{p} \partial y_{p}} \biggr)_{r \to 0} + 2 \eta_{mn} B_{m}
(\bec{\bf \nabla} \times {\bf B})_{n}
\nonumber\\
&& - 2 \hat \alpha_{mn} B_{m} B_{n} - \bec{\bf \nabla} \cdot {\bf
F} \;, \label{A7}
\end{eqnarray}
where we used an identity $ \langle {\bf h} \cdot (\bec{\bf \nabla}
\times {\bf h}) \rangle = (\partial^{2} \chi / \partial x_{p}
\partial y_{p})_{r \to 0} ,$ and $ {\bf r} = {\bf x} - {\bf y} .$
The second and third terms in Eq. (\ref{A7}) describes the sources of
the magnetic helicity. Therefore, the mean magnetic field $ {\bf B} ,$
the mean electric current $ \propto \bec{\bf \nabla} \times {\bf B} $ and the
hydrodynamic helicity  are the
sources of the magnetic helicity. The first term in  Eq. (\ref{A7})
determines the relaxation of the magnetic helicity with the
characteristic time $ T $ which depends on the molecular magnetic
diffusion $ \eta .$ This time is given by
\begin{eqnarray}
T^{-1} = {2 \eta \over \chi} \biggl({\partial^{2} \chi
\over \partial x_{p} \partial y_{p}} \biggr)_{r=0} \; .
\label{A7A}
\end{eqnarray}
The characteristic relaxation time
$ T $ of the magnetic helicity is $ T \sim
\tau_{0} {\rm Rm} ,$ i.e. it is much longer than the correlation time
$ \tau_{0} = l_{0} / u_{0} $ of the turbulent velocity field, where
$ u_{0} $ is the characteristic turbulent velocity in the maximum
scale of turbulent motions $ l_{0} .$
The last term in  Eq. (\ref{A7}) describes the
turbulent flux $ {\bf F} $ of the magnetic helicity which will be
calculated in Section III. Equation (\ref{A7}) in the case of
an isotropic turbulence coincides with that derived in \cite{KR82}
(see also \cite{GD94,KRR95}).

\section{The equation for the magnetic helicity tensor:
method of path integrals}

In this section we derive an equation for the magnetic helicity tensor.
To this purpose we use a method of path integrals
(see, e.g., \cite{RK97,ZRS90,EKR96,EKR98}).
This method allows us to derive the equation for the tensor $ \chi_{ij} =
\langle a_{i}({\bf x}) h_{j}({\bf y}) \rangle _{{\bf r} \to 0} :$
\begin{eqnarray}
{\partial \chi_{ij} \over \partial t} &=& - 2 \eta
\biggl({\partial^{2} \chi_{ij} \over \partial x_{p} \partial y_{p}}
\biggr)_{r=0} - 2 \tilde \eta_{np} \biggl({\partial^{2} \chi_{nj}
\over \partial x_{p} \partial y_{i}} \biggr)_{r=0}
\nonumber\\
&& + {\partial \over \partial R_{p}} (\varepsilon_{jpl}
\alpha_{ls}^{(v)} \chi_{is} - V_{p} \chi_{ij} + V_{l} \chi_{lj}
\delta_{ip})
\nonumber\\
&& + {\partial V_{j} \over \partial R_{p}} \chi_{ip} - {\partial
V_{p} \over \partial R_{i}} \chi_{pj} + 2 \alpha_{is}^{(v)} h_{sj}
\nonumber\\
&& - \alpha_{ks}^{(v)} h_{sk} \delta_{ij} + \varepsilon_{isp}
S_{p} h_{sj} + I_{ij} \; \label{L31}
\end{eqnarray}
(for details, see Appendix A), where $ {\bf R} = ( {\bf x} +  {\bf y}) /
2 ,$ and
\begin{eqnarray}
I_{ij} &=& \alpha_{is}^{(v)} B_{j} B_{s} - \alpha_{ks}^{(v)} B_{k}
B_{s} \delta_{ij} + \varepsilon_{ikl} \langle \tau u_{l} b \rangle
B_{k} B_{j}
\nonumber\\
&& + 2 \varepsilon_{kli} \tilde \eta_{lp}  B_{k} (\partial B_{j} /
\partial R_{p}) + J_{ij} \;,
\label{L32}
\end{eqnarray}
and $ h_{ij} = \langle h_{i}({\bf x}) h_{j}({\bf x}) \rangle ,$
and $ \tilde \eta_{ij} = \langle u_{i}({\bf x}) u_{j}({\bf x}) \rangle ,$
and $ S_{i} = \langle u_{i}({\bf x}) b({\bf x}) \rangle ,$
and $ \tilde \varphi_{j} = \langle \phi({\bf x}) h_{j}({\bf x})
\rangle ,$ and $ J_{ij} = \varepsilon_{lpj} \alpha_{ls}^{(v)} \langle
(\partial a_{i} / \partial x_{p}) h_{s} \rangle + \partial \tilde \varphi_{j}
/ \partial R_{i} - \langle (\partial h_{j} / \partial x_{i}) \phi
\rangle ,$ and $ b = \bec{\bf \nabla} \cdot {\bf u} .$
Equation (\ref{L31}) is derived for the case $ \bec{\bf \nabla}
\tilde \eta_{ij} = 0 .$
We use here the $ \delta $-correlated in time
random process to describe a turbulent velocity field. The results remain
valid also for the velocity field with a finite correlation time,
if the second-order correlation functions of the magnetic field and
the magnetic helicity vary
slowly in comparison with the correlation time of the turbulent
velocity field (see, e.g., \cite{ZRS90,DM84}). We also take into
account the dependence of the
momentum relaxation time on the scale of turbulent velocity field:
$ \tau ({\bf k}) = 2 \tau_{0} (k / k_{0})^{1-p} ,$ where $ p $ is the
exponent in spectrum of kinetic turbulent energy, $ k $ is the wave number,
$ k_{0} = l_{0}^{-1} .$
Equation for $ \chi = \chi_{pp} $ follows from Eq. (\ref{L31}):
\begin{eqnarray}
{\partial \chi \over \partial t} &=& - 2 \eta
\biggl({\partial^{2} \chi
\over \partial x_{p} \partial y_{p}} \biggr)_{r=0}
+ 2 \hat \eta_{mn} B_{m} (\bec{\bf \nabla} {\bf \times} {\bf B})_{n}
\nonumber\\
&& - 2 \alpha_{mn}^{(v)} B_{m} B_{n} + {\bf  \nabla}_{p}
[\varepsilon_{pmn} \chi_{ns} \alpha_{ms}^{(v)}
\nonumber\\
&& + V_{m} \chi_{mp} - (4/3) V_{p} \chi] \; \label{L36}
\end{eqnarray}
(see Appendix A), where hereafter $ {\bf  \nabla}_{p} =
\partial / \partial R_{p} ,$
and we used the gauge condition for the mean vector potential
$ \tilde \eta_{sp} {\bf  \nabla}_{p} A_{s} = 0 .$
For an isotropic turbulence $ (\tilde \eta_{mn} = \tilde \eta
\delta_{mn} / 3) $ the gauge condition is given by
$ \bec{\bf \nabla} {\bf \cdot} {\bf A} = 0 .$ The last term in Eq.
(\ref{L36}) describes the turbulent flux of the magnetic helicity
$ F_{p} = \varepsilon_{plt} \chi_{ts} \alpha_{ls}^{(v)}
+ V_{s} \chi_{sp} - (4/3) V_{p} \chi .$  The mean turbulent flux
of the magnetic helicity depends on the tensor of hydrodynamic
helicity $ \alpha_{ij}^{(v)}$ and the mean fluid velocity $ {\bf V}.$
Comparison of Eq. (\ref{L36}) [which was derived by the
path integral method] with Eq. (\ref{A7}) [which was obtained by the
simple consideration] shows that these two approaches arrive to the
similar equation after the change $ \alpha_{mn}^{(v)} \to
\alpha_{mn} .$ Note that the mean turbulent flux of the magnetic helicity
$ {\bf F} $ cannot be calculated by the simple consideration.

The tensor $ \chi_{ij} $ can be presented in the form $ \chi_{ij}
= \chi \delta_{ij} / 3 + \mu_{ij} ,$ where the anisotropic part of
the magnetic helicity tensor $ \mu_{ij} $ has the following
properties: $ \mu_{pp} = 0 .$ For the calculation of the second
spatial derivative $ (\partial^{2} \chi_{ij} / \partial x_{p}
\partial y_{p})_{r=0} $ we use  the tensor $ \chi_{ij}({\bf
k}^{(1)},{\bf k}^{(2)}) $ in $ {\bf k} $-space:
\begin{eqnarray}
\chi_{ij}({\bf  k}^{(1)},{\bf k}^{(2)}) = - 5 [ ( k_{pp}
\delta_{ij} - k_{ij}) (\chi_{\ast} / 5
\nonumber\\
- k_{mn} \mu_{nm} / 2 k_{pp}) - \mu_{im} k_{mj} - k_{im} \mu_{mj}
\nonumber\\
+ k_{pp} \mu_{ij} + k_{mn} \mu_{nm} \delta_{ij})] / 8 \pi k^{2} \;
, \label{L42}
\end{eqnarray}
where
\begin{eqnarray*}
\chi_{ij}({\bf x},{\bf y}) = \int \chi_{ij}({\bf k}^{(1)},{\bf
k}^{(2)})  \exp\{i( {\bf  k}^{(1)} {\bf x}
\nonumber\\
+ {\bf k}^{(2)} {\bf y})\} \,d^{3} k^{(1)} \, d^{3} k ^{(2)} ,
\end{eqnarray*}
and $ k_{ij} = k^{(2)}_{i} k^{(1)}_{j} .$ The tensor $
\chi_{ij}({\bf k}^{(1)},{\bf k}^{(2)}) $ satisfies the identities
$ k^{(1)}_{i} \chi_{ij}({\bf k}^{(1)},{\bf k}^{(2)}) = 0 $ and $
\chi_{ij}({\bf k}^{(1)},{\bf k}^{(2)}) k^{(2)}_{j} = 0 .$ These
identities correspond to the conditions $ \bec{\bf \nabla} \cdot
{\bf  a} = 0 $ and $ \bec{\bf \nabla} \cdot {\bf h} = 0 ,$
respectively. Using Eqs. (\ref{A7A}), (\ref{L42}) and (\ref{L43})
[see Appendix B] we rewrite Eq. (\ref{L36}) for the magnetic
helicity in the form
\begin{eqnarray}
\partial \chi / \partial t + \chi / T +
{\bf  \nabla}_{p} (V_{p} \chi)
+ 2 \alpha_{mn}^{(v)} B_{m} B_{n}
\nonumber\\
- 2 \hat \eta_{mn} B_{m} (\bec{\bf \nabla} {\bf \times} {\bf
B})_{n} = {\bf  \nabla}_{p} (\mu_{sf} \varepsilon_{fpl}
\alpha_{ls}^{(v)} + V_{s} \mu_{sp}) \; . \label{L44}
\end{eqnarray}
Equation (\ref{L44}) implies that the characteristic relaxation time
$ T $ of the isotropic part of the magnetic helicity tensor is $ T \sim
\tau_{0} {\rm Rm} ,$ i.e. it is much longer than the correlation time
$ \tau_{0} = l_{0} / u_{0} $ of the turbulent velocity field.
Equations (\ref{L31}) and (\ref{L44})
yield the equation for the tensor $ \mu_{ij} :$
\begin{eqnarray}
\eta^{\ast}_{jp} \mu_{pi} &+& 8 \eta^{\ast}_{ip} \mu_{pj} - 3
\mu_{ij} - 3 \delta_{ij} \eta^{\ast}_{pm} \mu_{mp} = (7/10) (3
\eta^{\ast}_{ij}
\nonumber\\
&& - \delta_{ij}) \chi + O(\tau_{0} / T) \;, \label{L45}
\end{eqnarray}
where $ \eta^{\ast}_{ij} = \tilde \eta_{ij} / \tilde \eta_{pp} .$
We neglected here small terms $ \sim (\tau_{0} / T) $ and $ \sim
\tau_{0} B^{2} .$ It follows from Eq. (\ref{L45}) that
the anisotropic part of the magnetic helicity tensor is determined
only by the turbulent diffusion tensor. Therefore, the
the characteristic time of evolution of the anisotropic part $
\mu_{ij}  $ of the magnetic helicity tensor is of the order of $
\tau_{0} ,$ i.e., it is very small. Solving Eq. (\ref{L45})
in the basis of the
eigenfunctions of the matrix $ \eta^{\ast}_{ij} $ we obtain
$ \mu_{ij} = 0 $ when $ i \not= j ,$ and $ \mu_{11} = \mu_{1} \chi ,$
and $ \mu_{22} = \mu_{2} \chi ,$ and $ \mu_{33} = -(\mu_{1} +
\mu_{2}) \chi ,$ where $ \eta^{\ast}_{ij} = 0 $ when $ i \not= j ,$
and $ \eta^{\ast}_{11} = \eta_{1} ,$ and $ \eta^{\ast}_{22} = \eta_{2} ,$
and $ \eta^{\ast}_{33} = 1 - (\eta_{1} + \eta_{2}) ,$ and
\begin{eqnarray*}
\mu_{1} = \biggl({7 \over 30} \biggr)
{\varepsilon_{2}^{2} - (\varepsilon_{1} - \varepsilon_{2})^{2} / 3
\over \varepsilon_{1} \varepsilon_{2} + (\varepsilon_{1} +
\varepsilon_{2})^{2} / 3} \;,
\\
\mu_{2} = \biggl({7 \over 30} \biggr)
{\varepsilon_{1}^{2} - (\varepsilon_{1} - \varepsilon_{2})^{2} / 3
\over \varepsilon_{1} \varepsilon_{2} + (\varepsilon_{1} +
\varepsilon_{2})^{2} / 3} \;,
\end{eqnarray*}
and $ \eta_{1} = 1/3 + \varepsilon_{1} $ and
$ \eta_{2} = 1/3 + \varepsilon_{2} .$
In the case of one preferential direction $
(\varepsilon_{1} = \varepsilon_{2} \equiv \varepsilon \not= 0) $ we obtain
$ \mu_{1} = \mu_{2} = 7 / 30 .$ When $ \varepsilon = 0 $
the anisotropic part of the magnetic helicity tensor
$ \mu_{ij} = 0 .$ In the case of one preferential direction
(say, in the direction $ {\bf e} ),$ Eqs. (\ref{L44}) and (\ref{L45})
yield
\begin{eqnarray}
\partial \chi / \partial t + \chi / T +
{\bf  \nabla}_{p} (V_{p}^{\rm eff} \chi)
+ 2 \alpha_{mn}^{(v)} B_{m} B_{n}
\nonumber\\
- 2 \hat \eta_{mn} B_{m} (\bec{\bf \nabla} {\bf \times} {\bf
B})_{n} = 0 \;, \label{L80}
\end{eqnarray}
where $ {\bf V}^{\rm eff} = 23 {\bf V} / 30 + 7 ({\bf e} \cdot {\bf
V}) {\bf e} / 10 - 7 ({\bf e} {\bf \times} {\bf D}) / 15 ,$ and
the vector $ D_{m} = \alpha_{mn}^{(v)} e_{n} .$
Equation (\ref{L80}) implies that even small anisotropy of turbulence
$ ( {\rm Rm}^{-1} \ll \varepsilon \ll 1) $ strongly modifies the
flux of the magnetic helicity.

For a weakly inhomogeneous turbulence the magnetic part of the
$ \alpha $-tensor is given by
\begin{eqnarray}
\alpha_{mn}^{(B)}({\bf r}=0) \sim {2 \chi \over 9 \eta_{T}
\mu_{0} \rho} \delta_{mn} \equiv \alpha^{(B)} \delta_{mn}
\;
\label{B6}
\end{eqnarray}
(see Appendix C), where $ \alpha^{(B)} = 2 \chi / (9 \eta_{T}
\mu_{0} \rho) $ and $ \chi = \chi({\bf R}) .$ This implies that the
tensor for the magnetic part of the $ \alpha $-effect
for weakly inhomogeneous turbulence is determined only by the isotropic
part of the magnetic helicity tensor. Thus, the evolutionary equation
for the magnetic part of the $ \alpha $-effect in this case is given by
\begin{eqnarray}
{\partial \alpha^{(B)} \over \partial t} + {\alpha^{(B)} \over T} +
{1 \over \rho} {\bf  \nabla}_{p} (V_{p}^{\rm eff} \alpha^{(B)}
\rho)
\nonumber\\
= - {4 \over 9 \eta_{T} \mu_{0} \rho} [\alpha_{mn}^{(v)} B_{m}
B_{n} -  \hat \eta_{mn} B_{m} (\bec{\bf \nabla} {\bf \times} {\bf
B})_{n}]  \;, \label{L81}
\end{eqnarray}
where we used Eqs. (\ref{L80}) and (\ref{B6}).

\section{Discussion}

We have shown here that an anisotropy of a fluid flow
strongly modifies the turbulent transport of the magnetic helicity.
In particular, even small anisotropy of turbulence
significantly changes the mean flux of the magnetic helicity. It is
given by $ {\bf F} = {\bf V}^{\rm eff} \chi .$ Indeed, if we consider,
e.g., a small anisotropy of turbulence: $ \varepsilon \sim
{\rm Rm}^{-\beta} $ (where $ \beta < 1 ),$ then the vector $ D_{m} =
\alpha_{pp}^{(v)} e_{m} / 3 + O({\rm Rm}^{-\beta}) .$ When the mean
velocity $ {\bf V} $ is normal to the vector $ {\bf e} $ (which is
typical for astrophysical applications) we obtain $ {\bf V}^{\rm eff}
\approx 23 {\bf V} / 30 .$ Therefore, a very small anisotropy $ \sim
{\rm Rm}^{-\beta} $ changes the mean flux of the magnetic helicity
to 25 percents. This result is associated with an existence of a small
parameter $ {\rm Rm}^{-1} $ which is the ratio of the relaxation times
of anisotropic and isotropic parts of the magnetic helicity tensor.
Note that the mean magnetic field is the main source of the magnetic
helicity. For zero mean magnetic field the magnetic helicity is very
small \cite{RK98}.

\begin{acknowledgments}
We have benefited from stimulating discussions with K.-H. R\"{a}dler.
This study was initiated by K.-H. R\"{a}dler during our visit in
Potsdam Institute of Astrophysics.
\end{acknowledgments}

\appendix
\section{Derivation of the equation for the magnetic
helicity tensor}

We use a method of path integrals (and modified Feynman-Kac formula)
(see, e.g., \cite{RK97,ZRS90,EKR96,EKR98}).
The solution of the
induction equation (\ref{A1}) with the
initial condition $ {\bf H}(t = t_0,{\bf x}) = {\bf H}_0( {\bf x}) $
is given by the Feynman-Kac formula
$ H_i (t,{\bf x}) = M \{ G_{ij} (t, t_{0}) H_{0j} [\bec{\bf \xi}
(t, t_{0})] \} ,$
where the function $ G_{ij} $ is determined by the equation
$ d G_{ij}(t_s, t_{0}) / ds = N_{ik} G_{kj} $ with the initial
condition $ G_{ij} = \delta _{ij} $ for $ t_{s}
= t_{0} .$ Here $ M \{ \cdot \} $ is a mathematical expectation
over the ensemble of Wiener paths, $ t_{s} = t + s ,$ and
$ N_{ik} = \partial v_i / \partial x_k - \delta _{ik} b,$
and the Wiener path, $ \bec{\bf \xi}_t \equiv  \bec{\bf \xi}(t, t_{0}) $
is given by
$ \bec{\bf \xi}_t = {\bf x} - \int_{t_{0}}^{t} {\bf v} (t_{s},{\bf \xi}_s )
\,ds + \sqrt{2 \eta} {\bf w}(t) ,$
where $ {\bf w}_t $ is a Wiener process.
This method allows us to get $ H_{i}(t + \Delta t,{\bf x}) :$
\begin{eqnarray}
H_{i}(t + \Delta t,{\bf x}) \simeq H_{i}(t,{\bf x}) + M \biggl\{
q_{i}({\bf x}) \Delta t
\nonumber\\
+ p_{i}({\bf x}) (\Delta t)^2 + \sqrt{2 \eta} Q_{in}({\bf x})
\int_{0}^{\Delta t} w_n \,d \sigma \} \; \label{L12}
\end{eqnarray}
(see Appendix in \cite{RK97}), where $ Q_{in} =  H_j \nabla_{n} N_{ij}
- (\nabla_{m} H_{i}) (\nabla_{n} v_{m}) ,$ and
\begin{eqnarray*}
q_{i} &=& H_m \nabla_{m} v_{i} - v_m \nabla_{m} H_{i} - b H_{i}
\\
&&+ \eta w_m w_n \nabla_{m} \nabla_{n} H_{i} (\Delta t)^{-1} \;,
\\
p_{i} &=& (1 / 2) H_{n}  [ \nabla_{m} (v_i \nabla_{n} v_{m} - v_m
\nabla_{n} v_{i}) - \nabla_{n} (b v_{i})
\\
&&+ \delta_{in} \nabla_{m} (b v_{m})] + \nabla_{m} H_{n} ( b v_{m}
\delta_{in} - v_{m} \nabla_{n} v_{i}
\\
&& + v_{k} \nabla_{k} v_{m} \delta_{in} / 2) + (1 / 2) v_{m}v_{n}
\nabla_{m} \nabla_{n} H_{i} \;,
\end{eqnarray*}
Now we use the following identity $ [\bec{\bf \nabla} {\bf \times}
(\hat \eta  \bec{\bf \nabla} {\bf \times} {\bf H}) ]_k =
{\bf \nabla}_{i} (H_{n} {\bf \nabla}_{n} \bar \eta_{ki}
- H_{i} {\bf \nabla}_{n} \bar \eta_{nk} - \bar \eta_{in} {\bf \nabla}_{n}
H_{k}) ,$ where $ \hat \eta \equiv \hat
\eta_{ij} = (\bar \eta_{pp} \delta_{ij} - \bar \eta_{ij} ) / 2 .$
This identity can be derived as follows. Consider the vector
$ E_{k} = {\bf \nabla}_{i} (H_{n} {\bf \nabla}_{n} \bar \eta_{ki})
= {\bf \nabla}_{i} {\bf \nabla}_{n} (H_{n} \bar \eta_{ki}) ,$ where
$ \bar \eta_{ki} $ is an arbitrary symmetrical tensor, and we use
the condition $ \bec{\bf \nabla} \cdot {\bf H} = 0 .$ Now we change
$ n \to i $ and $ i \to n .$ This yields
$ E_{k} = {\bf \nabla}_{n} {\bf \nabla}_{i} (H_{i} \bar \eta_{kn}) $
$ = {\bf \nabla}_{i} (H_{i} {\bf \nabla}_{n} \bar \eta_{kn} +
\bar \eta_{kn} {\bf \nabla}_{n} H_{i}) .$ Using this equation we calculate
the vector $ C_{k} = {\bf \nabla}_{i} (H_{n} {\bf \nabla}_{n} \bar \eta_{ki}
- H_{i} {\bf \nabla}_{n} \bar \eta_{nk} - \bar \eta_{in} {\bf \nabla}_{n}
H_{k}) = {\bf \nabla}_{i} (H_{i} {\bf \nabla}_{n} \bar \eta_{kn} +
\bar \eta_{kn} {\bf \nabla}_{n} H_{i} - H_{i} {\bf \nabla}_{n} \bar \eta_{nk}
- \bar \eta_{in} {\bf \nabla}_{n} H_{k}) = {\bf \nabla}_{i}
[(\bar \eta_{nk} \delta_{is} - \bar \eta_{in} \delta_{ks})
{\bf \nabla}_{n} H_{s}] .$
Now we introduce the tensor $ \hat \eta \equiv \hat \eta_{ij}
= (\bar \eta_{pp} \delta_{ij} - \bar \eta_{ij} ) / 2 .$
Using the identity $ \varepsilon_{kim} \varepsilon_{jns} \hat \eta_{mj} =
\bar \eta_{nk} \delta_{is} - \bar \eta_{in} \delta_{ks} $ we obtain
$ C_{k} = [\bec{\bf \nabla} {\bf \times} (\hat \eta \bec{\bf \nabla}
{\bf \times} {\bf H}) ]_k .$ Note that
the multiplication of the latter identity by $ \varepsilon_{kit}
\varepsilon_{fns} $ yields the definition of the tensor
$ \hat \eta_{ij} .$ Therefore, these calculations yield
the above identity.
Note that $ \bar \eta_{mn} $ is an arbitrary symmetrical tensor.
When $ \bar \eta_{mn} = W_{mn} $ (where $ \bec{\bf \nabla} W_{mn} = 0), $
these identities yield $ [\bec{\bf \nabla} {\bf \times} ( \hat W
\bec{\bf \nabla} {\bf \times} {\bf H}) ]_k = - W_{in}
{\bf \nabla}_{i} {\bf \nabla}_{n} H_{k} .$
We also use an identity $ v_i {\bf  \nabla}_{n} v_k = [\varepsilon_{ikp}
\bar{\alpha}_{pn}^{(v)} + \delta_{kn} \bar{S}_{i} - \delta_{in} \bar{S}_{k}
+ {\bf  \nabla}_{n} (v_{k} v_{k})] / 2 $ (see \cite{RK97}) where
$ \bar{\alpha}_{mn}^{(v)} = - (\varepsilon_{mji} v_i {\bf  \nabla}_{n} v_j
+ \varepsilon_{nji} v_i {\bf  \nabla}_{m} v_j) / 2 ,$ and
$ \bar{S}_{m} = v_{m} (\bec{\bf \nabla} \cdot {\bf v}) - {\bf  \nabla}_{n}
(v_{n} v_{m}) / 2 .$
Using these equations we obtain $ Q_{in} = \varepsilon_{itm}
{\bf  \nabla}_{t} (\varepsilon_{tps} H_{p} {\bf  \nabla}_{n} v_{s}) ,$
and
\begin{eqnarray}
q_{i} &=& \varepsilon_{itm} {\bf  \nabla}_{t} [ \varepsilon_{mls} v_l
H_s - \eta (w_{p} w_{p} \delta_{mn}
\nonumber\\
&& - w_{m} w_{n}) (\bec{\bf \nabla} {\bf \times} {\bf H})_n] / (2
\Delta t) \;,
\label{L20}\\
p_{i} &=& (-1 / 4) \varepsilon_{itm} {\bf  \nabla}_{t} [ \varepsilon_{mls}
H_s {\bf  \nabla}_{p} (v_l v_{p}) - 2 \bar{\alpha}_{mn}^{(v)} H_{n}
\nonumber\\
&& + (v_{p} v_{p} \delta_{mn} - v_{m} v_{n}) (\bec{\bf \nabla}
{\bf \times} {\bf H})_n] \; . \label{L21}
\end{eqnarray}
Equations (\ref{L12}), (\ref{L20})-(\ref{L21}) yield an equation
for the vector potential $ {\bf A}^{(t)} :$
\begin{eqnarray}
A^{(t)}_{i}(t + \Delta t,{\bf x}) \simeq A^{(t)}_{i}(t,{\bf x}) +
M \{ Q_{i}({\bf x}) \Delta t
\nonumber\\
+ P_{i}({\bf x}) (\Delta t)^2 + \sqrt{2 \eta} S_{in}({\bf x})
\int_{0}^{\Delta t} w_n \,d \sigma \}
\nonumber\\
+ \Delta t {\bf  \nabla}_{i} \varphi \;, \label{L23}
\end{eqnarray}
where $ {\bf H} = \bec{\bf \nabla} {\bf \times} {\bf A}^{(t)} ,$ and
$ S_{in} = \varepsilon_{ips} H_{p} {\bf  \nabla}_{n} v_{s} ,$ and
\begin{eqnarray}
Q_{i} &=& \varepsilon_{ifk} v_f H_k - \eta (w_{p} w_{p} \delta_{il}
\nonumber\\
&& - w_{i} w_{l}) (\bec{\bf \nabla} {\bf \times} {\bf H})_l / (2
\Delta t) \; .
\label{L24}\\
P_{i} &=& (-1/4) [\varepsilon_{ils} H_s {\bf  \nabla}_{p} (v_l v_{p}) -
2 \bar{\alpha}_{in}^{(v)} H_{n}
\nonumber\\
&& + (v_{p} v_{p} \delta_{in} - v_{i} v_{n}) (\bec{\bf \nabla}
{\bf \times} {\bf H})_n ] \;, \label{L25}
\end{eqnarray}
and $ \varphi $ is an arbitrary scalar function which depends on the
gauge condition.

Now we introduce a two-point correlation function $ \chi^{(xy)}_{ij} =
A_{ij} - A_{i}(t,{\bf x}) B_{j}(t,{\bf y}) ,$ where
$ A_{ij} = \langle A^{(t)}_{i}(t,{\bf x}) H_{j}(t,{\bf y})
\rangle ,$ and $ A^{(t)}_{i} =
A_{i} + a_{i} ,$ and $ H_{i} = B_{i} + h_{i} ,$ and $ {\bf A} =
\langle {\bf A}^{(t)} \rangle , \quad {\bf B} = \langle
{\bf H} \rangle ,$
where equations for the mean fields $ {\bf A} $ and $ {\bf B} $
are given by $ \partial B_{m} / \partial t = L^{(B)}_{mn} B_{n} ,$
and $ \partial A_{m} / \partial t = L^{(A)}_{mn} B_{n} + {\bf \nabla}_{m}
\tilde \Phi ,$ and
$ L^{(A)}_{sj}({\bf x}) = \varepsilon_{smj} V_{m} +
{\alpha}_{sj}^{(v)} - \hat \eta_{sm} \varepsilon_{mpj}
{\bf \nabla}_{p} ,$ and $ L^{(B)}_{ij}({\bf x}) = \varepsilon_{ips}
{\bf \nabla}_{p} L^{(A)}_{sj}({\bf x}) .$
Equations (\ref{L12}), (\ref{L20})-(\ref{L25}) yield
\begin{eqnarray}
\partial A_{ij} / \partial t &=& L^{(B)}_{js}({\bf y})
A_{is} + L^{(A)}_{ik}({\bf x}) H_{kj}
\nonumber\\
&& + N^{(xy)}_{ijks} H_{ks} + \phi_{ij} \;, \label{L27}
\end{eqnarray}
where $ \phi_{ij} = \langle ({\bf \nabla}_{i} \varphi({\bf x}))
H_{j}({\bf y}) \rangle ,$ and $ H_{ij} = \langle H_{i}({\bf x})
H_{j}({\bf y}) \rangle ,$ and
\begin{eqnarray*}
N^{(xy)}_{ijks} &=& {\alpha}^{(xy)}_{is} \delta_{kj} -
{\alpha}^{(xy)}_{ks} \delta_{ij} + \varepsilon_{ikf} S^{(xy)}_{f}
\delta_{js} - \varepsilon_{isk} S^{(xy)}_{j}
\nonumber\\
&+& \varepsilon_{ifk}
\biggl( {\partial \tilde \eta_{jf} \over \partial y_{s}} -
{\partial \tilde \eta_{pf} \over \partial y_{p}} \delta_{js} -
2 \delta_{js} \tilde \eta^{(xy)}_{fp} {\partial \over \partial
y_{p}} \biggr) \;,
\end{eqnarray*}
and
\begin{eqnarray*}
\alpha^{(xy)}_{mn} &=& - [ \varepsilon_{mji} \langle \tau u_i({\bf x})
{\partial u_{j}({\bf y}) \over \partial y_{n}} \rangle
\nonumber\\
&& + \varepsilon_{nji} \langle \tau u_i({\bf x}) {\partial
u_{j}({\bf y}) \over \partial y_{m}} \rangle] / 2 \;,
\nonumber\\
S^{(xy)}_{m} &=& \langle \tau u_{m}({\bf x}) b({\bf y}) \rangle
- (1 / 2) (\partial \tilde \eta_{mn} / \partial y_{n}) \;
\nonumber\\
\tilde \eta^{(xy)}_{mn} &=& \langle \tau u_{m}({\bf x}) u_{n}({\bf y})
\rangle \;, \quad \tilde \eta_{mn} = (\tilde \eta^{(xy)}_{mn} +
\tilde \eta^{(yx)}_{mn}) / 2 \;,
\end{eqnarray*}
and $ {\bf v} = {\bf V} + {\bf u} ,$ and $ {\bf V} =
\langle {\bf v} \rangle ,$ and $ b = \bec{\bf \nabla} \cdot {\bf u} .$
These tensors satisfy an identity
\begin{eqnarray*}
\langle \tau u_i({\bf x}) {\partial u_{k}({\bf y}) \over \partial
y_{n}} \rangle &=&  [\varepsilon_{ikm} \alpha^{(xy)}_{mn} +
\delta_{kn} S^{(xy)}_{i}
\nonumber\\
&& - \delta_{in} S^{(xy)}_{k} + {\partial \tilde \eta_{ki} \over
\partial y_{n}}] / 2 \;
\end{eqnarray*}
(see, e.g., \cite{R80,RK97}).
Similarly we introduce $ \alpha^{(yx)}_{mn} $ and $ S^{(yx)}_{m} :$
\begin{eqnarray*}
\alpha^{(yx)}_{mn} &=& - \biggl[ \varepsilon_{mji} \langle \tau
{\partial u_{i}({\bf x}) \over \partial x_{n}} u_j({\bf y})\rangle
\nonumber\\
&& + \varepsilon_{nji} \langle \tau {\partial u_{i}({\bf x}) \over
\partial x_{m}} u_j({\bf y}) \rangle \biggr] / 2 \;,
\nonumber\\
S^{(yx)}_{m} &=& \langle \tau u_{m}({\bf y}) b({\bf x}) \rangle
- (1 / 2) (\partial \tilde \eta_{mn} / \partial x_{n}) \;,
\end{eqnarray*}
which satisfy an identity
\begin{eqnarray*}
\langle \tau {\partial u_{i}({\bf x}) \over \partial x_{n}}
u_k({\bf y}) \rangle &=& [\varepsilon_{ikp} \alpha^{(yx)}_{pn} +
\delta_{kn} S^{(yx)}_{i}
\nonumber\\
&& - \delta_{in} S^{(yx)}_{k} + {\partial \tilde \eta_{ki} \over
\partial x_{n}}] / 2 \; .
\end{eqnarray*}
By means of Eq. (\ref{L27}) we derive an
equation for the tensor $ \chi^{(xy)}_{ij} $
\begin{eqnarray}
{\partial \chi^{(xy)}_{ij} \over \partial t} = L^{(B)}_{js}({\bf
y}) \chi^{(xy)}_{is} + L^{(A)}_{ik}({\bf x}) h_{kj} +
N^{(xy)}_{ijks} h_{ks}
\nonumber \\
+ N^{(xy)}_{ijks} B_{k}({\bf x}) B_{s}({\bf y}) + \langle ({\bf
\nabla}_{i} \phi({\bf x})) h_{j}({\bf y}) \rangle \;, \label{L30}
\end{eqnarray}
where $ h_{ij} = \langle h_{i}({\bf x}))
h_{j}({\bf y}) \rangle .$ Equation for the tensor $ \chi^{(yx)}_{ij} $
follows from Eq. (\ref{L30}) by the change $ {\bf x} \to {\bf y} $
and $ {\bf y} \to {\bf x} .$ Now we introduce a symmetrical tensor:
$ \chi_{ij} = (\chi^{(xy)}_{ij} + \chi^{(yx)}_{ij}) / 2 .$
Consider the case $ \bec{\bf \nabla} \tilde \eta_{mn} = 0 .$ Now we derive
equation for the tensor $ \chi_{ij}({\bf r}=0) $ using Eq. (\ref{L30}).
The result is given by (\ref{L31}).
For derivation of Eq. (\ref{L31}) we use the following identities:
\begin{widetext}
\begin{eqnarray}
N_{ijks} h_{ks} &=& \alpha_{is}^{(v)} h_{sj} - \alpha_{ks}^{(v)}
h_{sk} \delta_{ij} + \varepsilon_{isp} S_{p} h_{sj} + 2 \tilde
\eta_{mp} {\partial^{2} \chi_{mj} \over \partial x_{p}
\partial x_{i}} - 2 \tilde \eta_{pn} {\partial^{2} \chi_{ij} \over
\partial x_{p}\partial x_{n}} \;,
\label{L33} \\
L^{(A)}_{is}({\bf x}) h_{sj} &+& L^{(A)}_{is}({\bf y}) h_{js} =
\alpha_{is}^{(v)}({\bf x}) h_{sj} + \alpha_{is}^{(v)}({\bf y})
h_{sj} + \tilde \eta_{pn} \biggl( {\partial^{2} \chi^{(xy)}_{ij}
\over \partial x_{p} \partial x_{n}} + {\partial^{2}
\chi^{(yx)}_{ij} \over \partial y_{p} \partial y_{n} } \biggr)
\nonumber\\
&+& V_{s} \biggl({\partial \chi^{(xy)}_{sj} \over \partial x_{i}}
+ {\partial \chi^{(yx)}_{sj} \over \partial y_{i}} \biggr) - V_{s}
\biggl({\partial \chi^{(xy)}_{ij} \over \partial x_{s}} +
{\partial \chi^{(yx)}_{ij} \over \partial y_{s}} \biggr) \;,
\label{L34} \\
L^{(B)}_{js}({\bf x}) \chi^{(yx)}_{is} &+& L^{(B)}_{js}({\bf y})
\chi^{(xy)}_{is} = \varepsilon_{jpl} \biggl( {\partial \over
\partial y_{p}} (\alpha_{ls}^{(v)} \chi^{(xy)}_{ip}) + {\partial
\over \partial x_{p}} (\alpha_{ls}^{(v)} \chi^{(yx)}_{is}) \biggr)
- \biggl({\partial V_{p} \over \partial x_{p}} + V_{s} {\partial
\over \partial x_{s}} \biggr) \chi^{(yx)}_{ij}
\nonumber\\
&-& \biggl({\partial V_{p} \over \partial y_{p}} + V_{s} {\partial
\over \partial y_{s}} \biggr) \chi^{(xy)}_{ij} + \chi^{(yx)}_{ip}
{\partial V_{j} \over \partial x_{p}}  + \chi^{(xy)}_{ip}
{\partial V_{j} \over
\partial y_{p}} + \tilde \eta_{pn} \biggl( {\partial^{2}
\chi^{(xy)}_{ij} \over \partial y_{p} \partial y_{n}} +
{\partial^{2} \chi^{(yx)}_{ij} \over
\partial x_{p} \partial x_{n} } \biggr) \;, \label{L35}
\end{eqnarray}
\end{widetext}
\noindent
The tensor $ {\alpha}_{mn}^{(v)} = (\alpha^{(xy)}_{mn} +
\alpha^{(yx)}_{mn}) / 2 .$ We used here that
\begin{eqnarray*}
\biggl({\partial \chi^{(yx)}_{is} \over \partial x_{p}} +
{\partial \chi^{(xy)}_{is} \over \partial y_{p}}\biggr)_{r=0} = 2
\biggl({\bf \nabla}_{p} \chi_{is} - \langle {\partial a_{i} \over
\partial x_{p}} h_{s} \rangle\biggr) \; .
\end{eqnarray*}
The latter identity can be derived as follows.
\begin{eqnarray*}
\biggl({\partial \chi^{(yx)}_{is} \over \partial x_{p}} \biggr)_{r \to 0}
&=& \biggl(\langle {\partial h_{s}({\bf x}) \over \partial x_{p}}
a_{i}({\bf y}) \rangle \biggr)_{r \to 0}
\nonumber\\
&=& \biggl({\partial \over \partial x_{p}} \langle h_{s}({\bf x})
a_{i}({\bf y}) \rangle - \langle {\partial a_{i} \over
\partial x_{p}} h_{s} \rangle \biggr)_{r \to 0}
\\
&=& {\bf \nabla}_{p} \chi_{is} - \langle (\partial a_{i} /
\partial x_{p}) h_{s} \rangle \; .
\end{eqnarray*}
For the derivation of Eq. (\ref{L36}) we used the following identities:
\begin{eqnarray*}
\varepsilon_{ilk} \tilde \eta_{lp}  B_{k} {\bf \nabla}_{p} B_{i}
= - \hat \eta_{im} B_{i}
(\bec{\bf \nabla} {\bf \times} {\bf B})_{m} - B_{i} {\bf \nabla}_{i}
(\tilde \eta_{sp}  {\bf \nabla}_{p} A_{s}) \;,
\end{eqnarray*}
and $ \tilde \varphi_{p} = - V_{p} \chi / 3 + O(l_{0}^{2} /
l_{B}^{2}) ,$ where $ l_{B} $ is the characteristic scale of
the mean magnetic field variations, $ l_{0} $ is the maximum scale
of turbulent motions, and $ l_{0} \ll l_{B} .$

\section{The derivation of Eq. (\ref{L44})}

We use here the two-scale approach (see, e.g., \cite{RS75,KRA94}).
Indeed, let us consider, for  example,
a correlation function
\begin{eqnarray*}
\langle u_i ({\bf x} ) u_j ({\bf  y}) \rangle &=& \int \langle \,
u_i ({\bf  k}^{(1)}) u_j ({\bf k}^{(2)}) \, \rangle \, \exp\{i(
{\bf  k}^{(1)} \cdot {\bf x}
\\
&&+ {\bf  k}^{(2)} \cdot {\bf y})\} \,d^{3} k^{(1)} \, d^{3} k
^{(2)}
\\
&=& \int \tilde f_{ij}( {\bf r, K} ) \, \exp(i {\bf  K \cdot R})
\,d^{3} K
\\
&=& \int f_{ij}( {\bf k, R} ) \, \exp(i {\bf k \cdot r})  \,d^{3}
k \,,
\end{eqnarray*}
where $ \tilde f_{ij}({\bf  K ,  r} ) = \int \langle u_i ({\bf k}
+ {\bf  K} / 2 ) u_j( -{\bf k} + {\bf  K}  / 2 ) \rangle \exp(i
{\bf k \cdot r})  \,d^{3} k ,$ and $ f_{ij}({\bf k, R} ) = \int
\langle u_i ({\bf k} + {\bf  K} / 2 ) u_j( -{\bf k} + {\bf  K}  /
2 ) \rangle \exp{(i {\bf K \cdot R}) } \,d^{3} K ,$ and $ {\bf R}
= ( {\bf x} +  {\bf y}) / 2  , \quad {\bf r} =  {\bf y} - {\bf x},
\quad {\bf K} = {\bf k}^{(1)} + {\bf k}^{(2)},  \quad {\bf k} = (
{\bf k}^{(2)} -   {\bf k}^{(1)}) / 2 ,$ and $ {\bf R} $ and $ {\bf
K} $  correspond to the large scales, and $ {\bf r} $ and $ {\bf
k} $ describe the  small scales. Using Eq. (\ref{L42}) we obtain
\begin{eqnarray}
\eta^{\ast}_{mp} \biggl({\partial^{2} \chi_{ij} \over \partial
x_{m} \partial y_{p}} \biggr)_{r=0} = \tau_{0}^{-1} \biggl[ {\chi
\over 10} (3 \eta^{\ast}_{ij} - \delta_{ij})
\nonumber\\
+ {1 \over 7} (\eta^{\ast}_{jp} \mu_{pi} + 8 \eta^{\ast}_{ip}
\mu_{pj} - 3 \mu_{ij} - 3 \delta_{ij} \eta^{\ast}_{pm} \mu_{mp})
\biggr] \; , \label{L43}
\end{eqnarray}
where $ \chi({\bf r}=0) = \int \chi_{\ast} \exp(i {\bf K} \cdot {\bf R})
\,d^{3} {\bf K} \,d^{3} {\bf k} $ and $ \mu_{ij}({\bf r}=0) = \int
\mu_{ij}^{\ast} \exp(i {\bf K} \cdot {\bf R}) \,d^{3} {\bf K} \,d^{3} {\bf k} ,$
and $ \chi_{\ast} = \chi_{\ast}(k,{\bf K}) $ and $ \mu_{ij}^{\ast} =
\mu_{ij}^{\ast}(k,{\bf K}) .$ In order to obtain
Eq. (\ref{L43}) in $ {\bf r} $-space we used the transformations:
$ i{\bf  k}^{(1)} \to \partial / \partial {\bf x} $ and $ i{\bf  k}^{(2)} \to
\partial / \partial {\bf y} ,$ and we assumed a weak inhomogeneity of the
magnetic helicity, i.e., we neglected the terms $ \propto O({\bf K}) $
in Eq. (\ref{L42}). We also used the realisability
condition for the magnetic helicity (see, e.g., \cite{M78}),
i.e., we assumed that the spectral densities
$ \chi_{\ast} $ and $ \mu_{ij}^{\ast} \propto \chi $ are localized
in the vicinity of the maximum scale of turbulent motion  $ l_{0} .$
In order to derive Eq. (\ref{L43}) we used the following integrals:
\begin{eqnarray*}
&&Y_{ijmn} \equiv \int {k_{i} k_{j} k_{m} k_{n} \over k^{4}} \sin
\theta \,d \theta \,d \varphi = {4 \pi \over 15}
(\delta_{ij}\delta_{mn}
\\
&&+ \delta_{im}\delta_{nj} + \delta_{in}\delta_{mj}) \;,
\\
&&\int {k_{i} k_{j} k_{f} k_{s} k_{t} k_{r} \over k^{6}} \sin
\theta \,d \theta \,d \varphi = {1 \over 7} (Y_{fstr}\delta_{ij} +
Y_{jfsr}\delta_{it}
\\
&&+ Y_{ifsr}\delta_{jt} + Y_{jfst}\delta_{ir} +
Y_{ifst}\delta_{jr} + Y_{ijfs}\delta_{tr} - Y_{ijtr}\delta_{fs}) .
\end{eqnarray*}
Equations (\ref{L43}) and (\ref{L36}) allow us to obtain Eq. (\ref{L44}).

\section{The magnetic part of the $ \alpha $-effect for
weakly inhomogeneous turbulence}

In this Appendix we derive a formula for the magnetic part of
$ \alpha $-effect for weakly inhomogeneous turbulence.
We show that this tensor is determined by
the trace of the magnetic helicity tensor. The tensor $ \alpha_{mn}^{(B)} $
for the magnetic part of the $ \alpha $-effect is determined by Eq.
(\ref{A6}). Now we calculate
\begin{eqnarray*}
&& \varepsilon_{mji} \langle \tau h_i({\bf x}) {\bf  \nabla}_{n}
h_j({\bf y}) \rangle = - \varepsilon_{mji} \varepsilon_{lqi} \int
\tau({\bf k}^{(2)}) k_{l}^{(2)} k_{n}^{(1)}
\\
&& \times \langle a_{q}({\bf k}^{(2)}) h_j ({\bf k}^{(1)}) \rangle
\exp [i({\bf k}^{(1)} \cdot {\bf x}
\\
&& + {\bf k}^{(2)} \cdot {\bf y})] \,d {\bf k}^{(1)} \,d {\bf
k}^{(2)} = \int \tau({\bf k}^{(2)}) (k_{m}^{(2)} k_{n}^{(1)}
\chi_{pp}
\\
&& - k_{p}^{(2)} k_{n}^{(1)} \chi_{mp}) \exp [i({\bf k}^{(1)}
\cdot {\bf x} + {\bf k}^{(2)} \cdot {\bf y})] \,d {\bf k}^{(1)}
\,d {\bf k}^{(2)} \;,
\end{eqnarray*}
where $ \chi_{mn} = \langle a_m({\bf k}^{(2)}) h_n({\bf k}^{(1)}) \rangle .$
Since $ {\bf k}^{(2)} = {\bf k} + {\bf K} / 2 $ and
$ {\bf k}^{(1)} = - {\bf k} + {\bf K} / 2 ,$ we obtain
\begin{eqnarray}
&& \alpha_{mn}^{(B)}({\bf r}=0) = {1 \over \mu_{0} \rho} \int
\tau({\bf k}) [k_{m} k_{n} \chi_{pp} - K_{p} (k_{n} \chi_{mp}
\nonumber \\
&& + k_{m} \chi_{np}) - K_{m} K_{n} \chi_{pp} / 2 + K_{p} K_{n}
\chi_{mp}
\nonumber \\
&& + K_{p} K_{m} \chi_{np}] \exp [i{\bf K} \cdot {\bf R}] \,d {\bf
k} \,d {\bf K} \;, \label{B2}
\end{eqnarray}
where $ \rho $ is the fluid density, and $ \mu_{0} $ is the magnetic
permeability.
Equation (\ref{B2}) implies that the main contribution to the
tensor for the magnetic part of $ \alpha $-effect is from the trace for
the magnetic helicity tensor, i.e.,
$ \alpha_{mn}^{(B)}({\bf r}=0) \sim  \int \tau({\bf k}) k_{m} k_{n}
\chi_{pp}({\bf k},{\bf R}) \,d {\bf k} / \mu_{0} \rho .$
Now we assume that $ \chi_{pp}({\bf k},{\bf R}) \simeq \chi_{pp}(k,{\bf
R}) ,$ i.e., the trace of the magnetic helicity tensor in $ {\bf
k} $ space is isotropic (it is independent of the direction of $ {\bf
k} ).$ Therefore,
$ \alpha_{mn}^{(B)}({\bf r}=0) \sim  \delta_{mn} \int \tau(k) k^{2}
\chi_{pp}(k,{\bf R}) \,dk / (3 \mu_{0} \rho) ,$
where we used that $ \int (k_{m} k_{n} / k^{2}) \sin \theta \,d
\theta \,d \varphi = (4 \pi / 3) \delta_{mn} .$
The spectrum function of the magnetic helicity is given by
\begin{eqnarray*}
\chi(k,{\bf R}) &=& \chi({\bf R}) {c \over 4 \pi k^{2} k_0}
\biggl({k \over k_0}\biggr)^{-q} \,,
\\
c &=& (q-1) \biggl[ 1 - \biggl( {k_0 \over k_\chi} \biggr)^{q-1}
\biggr]^{-1} \;,
\end{eqnarray*}
where the wave number $ k $ is within interval $ k_{0} < k < k_\chi ,$ and
$ \chi({\bf R}) = \int \chi(k,{\bf R}) d {\bf k} ,$
and $ k_{0} = l_{0}^{-1} .$ The correlation
time is $ \tau(k) = 2 \tau_{0} (k / k_0)^{1-q} .$ The integration
in equation for $ \alpha_{mn}^{(B)}({\bf r}=0) $ yields
\begin{eqnarray}
\alpha_{mn}^{(B)}({\bf r}=0) &\sim& {\chi({\bf R}) (q-1) \over 9
(2-q) \eta_{T} \mu_{0} \rho} \biggl[\biggl( {k_\chi \over k_0}
\biggr)^{4-2q} - 1 \biggr]
\nonumber \\
&& \times \biggl[ 1 - \biggl( {k_0 \over k_\chi} \biggr)^{q-1}
\biggr]^{-1} \delta_{mn} \;, \label{B5}
\end{eqnarray}
The realisability condition causes $ k_\chi \simeq k_0 ,$ i.e., the
magnetic helicity is localized at the maximum scale of turbulent
motions (see, e.g., \cite{M78,ZRS83}). Therefore Eq. (\ref{B5}) yields
(\ref{B6}).

\end{document}